\documentstyle[12pt]{article}
\catcode`\@=11

\@addtoreset{equation}{section}
\def\theequation{\arabic{section}.\arabic{equation}}

\def\@normalsize{\@setsize\normalsize{15pt}\xiipt\@xiipt
\abovedisplayskip 14pt plus3pt minus3pt%
\belowdisplayskip \abovedisplayskip
\abovedisplayshortskip  \z@ plus3pt%
\belowdisplayshortskip  7pt plus3.5pt minus0pt}
\def\small{\@setsize\small{13.6pt}\xipt\@xipt
\abovedisplayskip 13pt plus3pt minus3pt%
\belowdisplayskip \abovedisplayskip
\abovedisplayshortskip  \z@ plus3pt%
\belowdisplayshortskip  7pt plus3.5pt minus0pt
\def\@listi{\parsep 4.5pt plus 2pt minus 1pt
            \itemsep \parsep
            \topsep 9pt plus 3pt minus 3pt}}

\def\underline#1{\relax\ifmmode\@@underline#1\else
        $\@@underline{\hbox{#1}}$\relax\fi}
\@twosidetrue
\relax

\catcode`@=12

\catcode`\@=11

\def\section{\@startsection{section}{1}{\z@}{3.5ex plus 1ex minus
   .2ex}{2.3ex plus .2ex}{\large\bf}}
\def\thesection{\arabic{section}.}

\def\ps@headings{\def\@oddfoot{}\def\@evenfoot{}
\def\@oddhead{\hbox{}\hfill
        \makebox[.5\textwidth]{\raggedright\ignorespaces --\thepage{}--
        \hfill }}
\def\@evenhead{\@oddhead}
\def\subsectionmark##1{\markboth{##1}{}}
}

\ps@headings

\catcode`\@=12

\relax

\def\figcap{\section*{Figure Captions\markboth
        {FIGURECAPTIONS}{FIGURECAPTIONS}}\list
        {Fig. \arabic{enumi}:\hfill}{\settowidth\labelwidth{Fig. 999:}
        \leftmargin\labelwidth
        \advance\leftmargin\labelsep\usecounter{enumi}}}
 \relax
\def\tablecap{\section*{Table Captions\markboth
        {TABLECAPTIONS}{TABLECAPTIONS}}\list
        {Table \arabic{enumi}:\hfill}{\settowidth\labelwidth{Table 999:}
        \leftmargin\labelwidth
        \advance\leftmargin\labelsep\usecounter{enumi}}}
 \relax
\def\reflist{\section*{References\markboth
        {REFLIST}{REFLIST}}\list
        {[\arabic{enumi}]\hfill}{\settowidth\labelwidth{[999]}
        \leftmargin\labelwidth
        \advance\leftmargin\labelsep\usecounter{enumi}}}
 \relax

\catcode`\@=11

\def\ps@headings{\def\@oddfoot{}\def\@evenfoot{}
\def\@oddhead{\hbox{}\hfill
        \makebox[.5\textwidth]{\raggedright\ignorespaces --\thepage{}--
        \hfill }}
\def\@evenhead{\@oddhead}
\def\subsectionmark##1{\markboth{##1}{}}
}

\ps@headings

\relax

\def\firstpage#1#2#3#4#5#6{
\begin{document}
\begin{titlepage}
\nopagebreak
\title{\begin{flushright}
        \vspace*{-1.5in}
        {\normalsize NUB--#1\\[-3mm]
        #2\\[-9mm]CPTH--A258.0793}\\[6mm]
\end{flushright}
\vfill
{\large \bf #3}}
\author{\large #4 \\[1cm] #5}
\maketitle
\vfill
\nopagebreak
\begin{abstract}
{\noindent #6}
\end{abstract}
\vfill
\begin{flushleft}
\rule{16.1cm}{0.2mm}\\[-3mm]
$^{\star}${\small Work supported in part by\vspace{-4mm}
the National Science Foundation under grants PHY--91--07809
and PHY--93--06906, in part by the EEC contracts \vspace{-4mm}
SC1--915053 and SC1--CT92--0792, and in part by CNRS--NSF
grant INT--92--16146.}\\
July 1993
\end{flushleft}
\thispagestyle{empty}
\end{titlepage}}
\newcommand{\NIJ}{{\cal N}_{IJ}}
\newcommand{\N}{{\cal N}}
\newcommand{\dal}{\raisebox{0.085cm}
{\fbox{\rule{0cm}{0.07cm}\,}}}
\newcommand{\dt}{\partial_{\langle T\rangle}}
\newcommand{\dtbar}{\partial_{\langle\fracline{T}\rangle}}
\newcommand{\al}{\alpha^{\prime}}
\newcommand{\mst}{M_{\scriptscriptstyle \!S}}
\newcommand{\mpl}{M_{\scriptscriptstyle \!P}}
\newcommand{\dv}{\int{\rm d}^4x\sqrt{g}}
\newcommand{\lv}{\left\langle}
\newcommand{\rv}{\right\rangle}
\newcommand{\ph}{\varphi}
\newcommand{\sbar}{\,\fracline{\! S}}
\newcommand{\xbar}{\,\fracline{\! X}}
\newcommand{\barz}{\,\fracline{\! Z}}
\newcommand{\zbar}{\bar{z}}
\newcommand{\dbar}{\,\fracline{\!\partial}}
\newcommand{\tbar}{\fracline{T}}
\newcommand{\psibar}{\fracline{\Psi}}
\newcommand{\ybar}{\fracline{Y}}
\newcommand{\z}{\zeta}
\newcommand{\zb}{\bar{\zeta}}
\newcommand{\phb}{\fracline{\varphi}}
\newcommand{\cm}{Commun.\ Math.\ Phys.~}
\newcommand{\pr}{Phys.\ Rev.\ D~}
\newcommand{\pl}{Phys.\ Lett.\ B~}
\newcommand{\np}{Nucl.\ Phys.\ B~}
\newcommand{\e}{{\rm e}}
\newcommand{\gsi}{\,\raisebox{-0.13cm}{$\stackrel{\textstyle
>}{\textstyle\sim}$}\,}
\newcommand{\lsi}{\,\raisebox{-0.13cm}{$\stackrel{\textstyle
<}{\textstyle\sim}$}\,}
\date{}
\firstpage{3071}{IC/93/202}
{\large\sc Topological Amplitudes in String
Theory$^\star$}
{I. Antoniadis$^{\,a}$, E. Gava$^{b,c}$,
K.S. Narain$^{ c}$ $\,$and$\,$
T.R. Taylor$^{\,d}$}
{\normalsize\sl
$^a$Centre de Physique Th\'eorique, Ecole Polytechnique,
F-91128 Palaiseau, France\\[-3mm]
\normalsize\sl (Laboratoire Propre du CNRS UPR A.0014)\\
\normalsize\sl
$^b$Instituto Nazionale di Fisica Nucleare, sez.\ di Trieste,
Italy\\
\normalsize\sl $^c$International Centre for Theoretical Physics,
I-34100 Trieste, Italy\\
\normalsize\sl $^d$Department of Physics, Northeastern
University, Boston, MA 02115, U.S.A.}
{We show that certain type II string amplitudes at genus $g$ are given by
the topological partition function $F_g$ discussed recently by Bershadsky,
Cecotti, Ooguri and Vafa. These amplitudes give rise to a term in the
four-dimensional
effective action of the form $\sum_g F_g W^{2g}$, where $W$ is the
chiral superfield of $N=2$ supergravitational multiplet. The
holomorphic anomaly of $F_g$ is related to non-localities of the
effective action due to the propagation of massless states. This result
generalizes the holomorphic anomaly of the one loop case which is known
to lead to non-harmonic gravitational couplings.}

\setcounter{section}{0}
\section{Introduction}

In the recent years, there has been some progress in understanding the
string loop corrections in the effective low energy supergravity theory
in four dimensions.
This study was initiated by Dixon, Kaplunovsky and Louis
\cite{DKL} who determined the dependence of the one-loop
threshold corrections to
gauge couplings on the untwisted moduli in orbifold models.
Their computation was later
generalized to arbitrary supersymmetric compactifications and it was
shown \cite{AGN} that the one-loop gauge couplings are given by a quantity
which is very closely related to the new supersymmetric index \cite{CFIV}
of the internal $N=2$ superconformal theory. Moreover, the
non-harmonicity of the group-dependent part of gauge couplings, which
translates into a differential equation with respect to the moduli, was
shown to correctly reproduce the results of the low energy effective field
theory \cite{ft}. On the other hand, the differential equation has the
same form as the one obeyed by the new supersymmetric index. Subsequently,
these computations were extended to obtain the one-loop corrections to the
K\"ahler metric and Yukawa couplings \cite{AGNT}, which complete the form
of the effective Lagrangian up to two derivatives. All of these
quantities were also related to the new supersymmetric index.

In a recent work, Bershadsky, Cecotti, Ooguri and Vafa
showed that the new supersymmetric
index for the internal $N=2$ superconformal theory can be in fact
understood as the one-loop topological partition function $F_1$ obtained
by twisting the internal theory \cite{BCOV}. They also considered the
higher genus partition functions $F_g$ ($g$ being the genus) of the
topological theory. Since $N=2$ antichiral fields are BRS exact one would
naively expect $F_g$'s to be holomorphic. However, just as in the case of
$F_1$, at higher genera also there is a holomorphic anomaly which leads
to recursion relations between $F_g$'s. Furthermore these recursion
relations could be summarized in the form of a master equation for a
generating function. Since $F_1$ appears also in the string one-loop
effective action, a natural question arises whether higher $F_g$'s also
have something to do with string amplitudes. If so the existence of a
master equation for these higher loop effective action terms could provide
a powerful tool to obtain some possible non-perturbative information. The
purpose of this paper is to establish such a connection between certain
string amplitudes and $F_g$'s.

Although, as mentioned earlier, heterotic string one-loop computation
gives rise to $F_1$ when difference of the gauge couplings for $E_6$ and
$E_8$ is considered, $F_1$ appears more directly in the context of type II
strings. Indeed, by following the same steps as in the heterotic
case, one can easily show that  $F_1$ corresponds to the one-loop correction to
gravitational coupling of the form $R^2$. One might expect that since type
II string is left-right symmetric, it may be more easy to establish the
correspondence with the topological theory which also is left-right
symmetric. Therefore in the following we restrict ourselves to the type
II case.

The question arises what kind of amplitudes in type II string could
possibly be related to topological theory. Since the topological theory is
obtained by twisting the internal $N=2$ theory \cite{Top}, one might try to
consider amplitudes which effectively would introduce enough background
charges for the internal $U(1)$ of the $N=2$ algebra to twist the theory.
One way to satisfy this requirement, as proposed by Bershadsky et al.\
\cite{bp}, is to consider amplitudes involving $2g\! -\!2$ Ramond-Ramond
fields carrying $U(1)$ charges $3/2$ on a genus $g$ Riemann surface.
Type II strings lead to $N=2$ space-time supersymmetry and the $N=2$
graviphoton vertices are exactly of this type. Therefore one is naturally
led to consider amplitudes involving $2g\! -\!2$ graviphotons. This suggests
a possible term in the effective action of the form $W^{2g}$, where $W$
is the chiral superfield of  $N=2$ supergravitational multiplet whose first
component is the graviphoton field strength $T_{\mu\nu}$ \cite{chiraln2}.
Its expansion in components
contains a term of the form $R^2 T^{2g-2}$. In this paper we
show that amplitudes
corresponding to such a term are in fact given by $F_g$ which therefore
has the interpretation of being the moduli-dependent coupling associated to
$W^{2g}$. Similar results have also been obtained by Bershadsky et al.\
\cite{bf}.

Since $W$ is a chiral superfield, a local term in the action of the form
$F_g W^{2g}$ would require $F_g$ to be holomorphic in moduli. However,
just as in the one-loop case, $F_g$ becomes non-holomorphic due to the
propagation of massless particles which leads to non-locality in the
effective action. It is remarkable that this information is provided by
the holomorphic anomaly of the topological theory.

The paper is organized as follows: In Section 2, we review some of the
basic features of type II strings. In Section 3, we present the
computation of the genus $g$ amplitude involving two gravitons and
$2g\! -\!2$ graviphotons for orbifolds, and show that it reproduces the
topological partition function $F_g$. In Section 4, we show that this
result generalizes to arbitrary Calabi-Yau compactifications. In Section
5, we discuss connection with the effective $N=2$ supergravity theory.
Section 6 contains conclusions.
In the Appendix, we give details of the derivation
of the effective action terms corresponding
to the amplitudes computed in Sections 3 and 4.

\section{Vertex operators in type II superstrings}

In this section we will review some of the features of type II
strings that we will be using in the following parts of the paper.
We will be considering type II theories giving rise to $N=2$
supergravity in four dimensions, i.e. one supersymmetry from
the left sector and one from the right. This means that
the compactification  from ten to four dimensions
is performed via a $(2,2)$ superconformal
field theory (SCFT) with $c=9$. We will consider in detail
orbifolds, but also general Calabi-Yau spaces.

Let us start to discuss the massless spectrum
of this type of models from the universal sector,
which is independent of the particular compactification
sheme adopted, and is constituted by states
corresponding to the identity operator in the $N=2$ internal
theory and those which are obtained from them through spectral
flow, i.e. space-time supersymmetry. In this case we have two
multiplets: the gravitational one, whose bosonic components
include the graviton and a vector field, the
``graviphoton", which enters in the supersymmetry transformation
of the gravitinos. The corresponding field
strength is denoted by $T_{\mu\nu}$.
Then there is the universal hypermultiplet,
which contains the dilaton and the antisymmetric tensor, along
with two other Ramond-Ramond (R-R) scalars.

There are in general additional vector multiplets and
hypermultiplets: to describe their structure one has to distinguish
between type IIA and type IIB theories. In type IIA case, the
complex scalars belonging to the vector
multiplets are the $(1,1)$ moduli
of the internal $N=2$ theory,
corresponding to states with $U(1)$ charges
$(q=-\bar q=1)$. The corresponding R-R vector fields are obtained
by spectral flow. On the other hand, the (NS-NS) scalars within the
hypermultiplets corresponds to the $(1,2)$ moduli, i.e. those
with $(q=\bar q=1)$. The other R-R scalars are similarly obtained
by spectral flow.
In type IIB case the assignement is reversed \cite{Sei,CFG}.

Our analysis in the following will be, for simplicity,
mainly restricted to
the gravitational sector of type IIA (and IIB)
theories, that is we will
study amplitudes involving gravitons and graviphotons.
Let us then recall the expressions for the corresponding
vertex operators. The graviton vertex operator, in the $0$ ghost
picture, is:\footnote{We adopt the following notation:
right-moving fields are written with a tilde, $U(1)$ charge
conjugated with a bar.}
\begin{equation}
V_g^{(0)}(p,h) ~=~ :\!h_{\mu\nu} (\partial X^{\mu}+ip\cdot\psi
 \psi^{\mu})({\bar{\partial}} X^{\nu}
+ip\cdot{\widetilde\psi}{\widetilde\psi}^{\nu})
e^{ip\cdot \!X}\!:~,
\label{vg}
\end{equation}
Here $h_{\mu\nu}$ is symmetric, traceless and obeys $p^{\mu}h_{\mu\nu}=0$.
As for the graviphoton, it is a R-R
state and its vertex operator, in the $-1/2$ picture, is given
by \cite{DKV,CFG}:
\begin{equation}
V_T^{(-1/2)}(p,\epsilon) ~=~ :e^{-1/2(\phi+\widetilde\phi)}
p_{\nu}\epsilon_{\mu}[S^{\alpha}(\sigma^{\mu\nu})_{\alpha}^{\beta}
{\widetilde S}_{\beta}\Sigma(z,\zbar)+ S_{\dot{\alpha}}({\bar\sigma}^{\mu
\nu})^{\dot{\alpha}}_{\dot{\beta}}{\widetilde S}^{\dot{\beta}}{\bar\Sigma}
(z,\zbar)]e^{ip\cdot \!X}\!:~,
\label{vgf}
\end{equation}
where the polarization $\epsilon_{\mu}$ obeys $p\cdot\epsilon=0$,
$\phi$ is the free scalar bosonizing the superghost system,
$S_{\alpha}$, $S^{\dot{\beta}}$
are space-time spin fields of opposite helicities and $\Sigma$
is a field in the internal $N=2$ sector of $U(1)$ charges $(q=-\bar q
=3/2)$ in the type IIA case and $(q=\bar q=3/2)$ in type IIB case.
Bosonizing in
the standard way the internal $U(1)$ current $J$ ($\widetilde J$)
by the free boson $H$ ($\widetilde H$), $J = i\sqrt{3}\partial H$ and
similarly for $\widetilde J$, one can explicitly write:
\begin{equation}
\Sigma(z,\zbar)~=~{\rm exp}[i\frac{\sqrt{3}}{2}(H(z) \mp{\widetilde
H}(\zbar))],
\label{sigma}
\end{equation}
where the upper (lower) sign case corresponds to type IIA
(type IIB) theories.
For the other gauge fields one has the same structure as in
eq.(\ref{vgf}), but now with $\Sigma$ replaced by charge $(-1/2,1/2)$
fields in case IIA and $(-1/2,-1/2)$ in case IIB.

We will be interested in gauge fields with definite duality (helicity)
properties: this means  we will choose
the polarization $\epsilon_{\mu}$ corresponding to an anti-self-dual
$T_{\mu\nu}=T^{-}_{\mu\nu}$. As a result only the first
term in (\ref{vgf}) will enter in our amplitudes. Analogously
we will consider anti-self-dual gravitons by choosing
$h_{\mu\nu}$ corresponding to an anti-self-dual Riemann tensor
$R_{\mu\nu\rho\sigma}=
R^{-}_{\mu\nu\rho\sigma}$.
Throughout this paper, we adopt a notation in which all
antisymmetric tensors (without explicit $-$ or $+$ superscripts)
are automatically assumed to be anti-self-dual.

Let us now briefly consider hypermultiplets, starting from the universal
one. The dilaton and antisymmetric tensor vertex operators are
of course obtained from (\ref{vg}) with the appropriate choice of
$h_{\mu\nu}$. The corresponding R-R scalar vertex operator is
given by:
\begin{equation}
V^{(-1/2)}_{Z}(p)~=~:e^{{-1/2}(\phi+\widetilde\phi)}p_{\mu}
S^{\alpha}(\sigma^{\mu})_{\alpha\dot{\beta}}{\widetilde S}^{\dot{\beta}}
\Phi(z,\bar z) e^{ip\cdot \!X}\!:~,
\label{rs}
\end{equation}
along with its conjugate, and $\Phi$ is the same as $\Sigma$ given in
(\ref{sigma}) with $\mp$ replaced by $\pm$. The R-R members of the other
hypermultiplets have a similar structure,
with the appropriate modification of the internal part, where fields
of charges $(|q|=|\bar q|=1/2)$ appear.

As a final remark, the basis of fields one has naturally
in the string formulation, which corresponds to the above choice
of vertex operators, in general does not coincide with the one
usually adopted in the $N=2$ supergravity literature \cite{dWL}. We will
comment on the relation between the two basis in section $5$.

\section{String computation for orbifolds}

In the case of orbifolds, the internal $N=2$ SCFT is realized
in terms of free bosons and fermions. We consider for simplicity
orbifolds realized in terms of $3$ complex bosons $X_{I}$ and
left- (right-) moving fermions $\psi_{I}$ ($\widetilde{\psi}_{I}$),
with $I=3,4,5$. Let $h$ be an element of the orbifold group
defined by $h=\{h_{I}\}$, and its action on $X_I$ is
$X_{I}\rightarrow e^{2\pi i h_{I}}X_I$ and similarly for
$\psi_I$ and $\widetilde{\psi}_{I}$. Space-time supersymmetry implies that
one can always choose the $h_I$'s to satisfy the condition:
\begin{equation}
\sum_{I} h_{I}~=~0\ .
\label{susy}
\end{equation}
On a genus $g$ Riemann surface we must associate
to each homology cycle $a_{i},b_{i}$, for $i=1,\cdots,g$, an element
of the orbifold group. In the following we shall denote by
$\{h\}=\{\{h_I\}\}$ the set of all twists along different cycles.
One can bosonize the complex fermions
\begin{equation}
\psi_{I}=e^{i \phi_I}~,~~~~~~~~~~~~~\bar{\psi}_{I}=e^{-i\phi_I}~,
\label{boso}
\end{equation}
and similarly for $\widetilde{\psi}$,$\,\bar{\!\widetilde{\psi}}$.
The previously introduced free boson $H$ can be expressed
in terms of the $\phi_{I}$'s as:
$\sqrt{3}H=\sum_{I=3}^{5}{\phi}_{I}$, and similarly for the right-moving
one. We can also bosonize the space-time fermionic coordinates by
arranging them into complex left-(right-) moving fermions
$\psi_{1,2}=e^{i\phi_{1,2}}$,
$\widetilde{\psi}_{1,2}=e^{i\widetilde{\phi}_{1,2}}$. As already
mentioned, we also bosonize the $\beta ,\gamma$ system in terms of a free
boson $\phi$ and the $\eta ,\xi$ system as usual  \cite{FMS}.

Consider now, at $g$-th string-loop order,
the following amplitude:
\begin{equation}
A_{g}~=~\langle T^{2g-2}R^{2}\rangle_{g}
{}~.\label{ampli}
\end{equation}
As we explained in section 2, the vertex operators for anti-self-dual
graviphotons contain, in the $-1/2$
ghost picture, a combination of spin fields $S_{\alpha}$, ${\widetilde S}_
{\beta}$, whose precise form is governed by kinematics. Here:
\begin{equation}
S_{1} ~=~ {\rm exp}[\frac{i}{2}(\phi_1+\phi_2)]~,
\label{s1}\vspace{1mm}
\end{equation}
\begin{equation}
S_{2}~=~ {\rm exp}[-\frac{i}{2}(\phi_1+\phi_2)]~,
\label{s2}
\end{equation}
and similarly for the right-moving parts $\widetilde S$, where $\phi$'s
are replaced by $\widetilde\phi$'s.
The anti-self-dual part of the graviton vertex in the $0$ ghost picture
contains the following left-moving fermionic combinations:
$\psi_1\psi_2$, ${\bar\psi}_1{\bar\psi}_2$ and $\psi_1{\bar\psi}_1
+\psi_2{\bar\psi}_2$, together with similar terms for the
right-moving fermionic parts, depending on the kinematics.

We will see below that the purely bosonic part of the graviton
vertices, $\partial X^{\mu}\bar{\partial}X^{\nu}$, does not contribute
to the amplitude under consideration.
To make the calculation transparent we choose a kinematical
configuration such that $g\!-\!1$ of the $T$'s appear with
$S_1{\widetilde S}_1$ and the remaining $g\!-\! 1$ $T$'s with
$S_2{\widetilde S}_2$ . Furthermore we take one graviton vertex to be
$\psi_1\psi_2{\widetilde\psi}_1{\widetilde\psi}_2$, and the second one
as ${\bar\psi}_1{\bar\psi}_2
{\,\bar{\!\widetilde\psi}}_1{\,\bar{\!\widetilde\psi}}_2$.

Also we change $g\!-\!1$ of the $T$'s to $+1/2$ ghost picture:
this is done by inserting $g\! -\! 1$ picture-changing operators
$\delta(\beta) T_F$, $T_F$ being the total
worldsheet supercurrent. On a
genus $g$ Riemann surface there are also $2g\! -\!2$ additional
insertions of picture-changing operators, as a result of the
integration over supermoduli. Thus the total number of insertions
of picture-changing operators is $3g\! -\!3$.

The amplitude then becomes:\footnote{In the following we will omit
overall $g$-independent constants, as well as normalizations of the form
$a^g$ which can be absorbed by
constant rescalings of the fields.}
\begin{eqnarray}
A_g~&=&~\langle\prod_{i=1}^{g-1} e^{-1/2(\phi+\widetilde{\phi})}
S_1{\widetilde S}_1\Sigma(x_i,\bar{x}_i)
\prod_{i=1}^{g-1} e^{-1/2(\phi+\widetilde{\phi})}
S_2{\widetilde S}_2\Sigma(y_i,\bar{y}_i)\nonumber\\ & &
\psi_1\psi_2{\widetilde\psi}_1{\widetilde\psi}_2(z)
{\bar\psi}_1{\bar\psi}_2{\,\bar{\!\widetilde\psi}}_1
{\,\bar{\!\widetilde\psi}}_2(w) \prod_{i=1}^{3g-3} e^{\phi} T_{F}(z_i)
e^{\widetilde\phi}{\widetilde T}_{F}(\bar{z}'_i){\rangle}_g~.
\label{ampl}
\end{eqnarray}
Since all the $T$'s carry charge $+\frac{1}{2}$ for
$\phi_3$, $\phi_4$, $\phi_5$, $g\! -\! 1$ of the $T_F$'s must
provide $-1$ charge each for $\phi_3$, $\phi_4$ and $\phi_5$
respectively. We first discuss the left-moving part of the
correlation function. We have the following contributions \cite{vv}:

a) From the space-time part:
\begin{eqnarray}
A_{g}^{(st)}&=&\frac{\theta_{s}^2(\frac{1}{2}\sum_{i=1}^{g-1}(x_{i}-
y_{i})+z-w)}{Z_1} {\frac{\prod_{i<j} E^{1/2}(x_i,x_j) E^{1/2}(y_i,y_j)}
{\prod_{i<j} E^{1/2}(x_i,y_j)}}\nonumber\\ & &
{\frac{\prod_{i} E(x_i,z) E(y_i,w)}
{\prod_{i} E(x_i,w) E(y_i,z)}}{\frac{1}{E^{2}(z,w)}}~,\label{spt}
\end{eqnarray}
where $\theta_s$ denotes the genus $g$ $\theta$-function of
spin structure $s$, $E$ is the prime form and $Z_1$ is the
chiral determinant of the $(1,0)$ system.

b) From the superghost part:
\begin{eqnarray}
A_{g}^{(sgh)}&=&\frac{Z^{1/2}_{1}}
{\theta_{s}(\frac{1}{2}\sum_{i}^{g-1}(x_i+y_i)
-\sum_{a}^{3g-3}z_a+2\Delta)}\nonumber\\ & &
\frac{\prod_{i,a}E^{1/2}(x_i,z_a) E^{1/2}(y_i,z_a)}
{\prod_{i<j}E^{1/4}(x_i,x_j) E^{1/4}(y_i,y_j) \prod_{i,j}
E^{1/4}(x_i,y_j)}\nonumber\\ & &
\frac{\prod_{i}\sigma(x_i)\sigma(y_j)}
{\prod_{a<b}^{3g-3} E(z_a,z_b)\prod_{a}{\sigma}^{2}(z_a)}
{}~,\label{sgh}
\end{eqnarray}
where $\Delta$ is the Riemann $\theta$-constant, which represents
the degree $g\!-\!1$ divisor of a half differential associated with
the preferred spin structure for a given marking of the Riemann
surface. $\sigma(x)$ is a $g/2$-differential with no zeros or
poles, transforming in a quasiperiodic way under $a_i$ or $b_i$
monodromies.

c) From the internal part: as mentioned above, $g\!-\! 1$ of the
$T_F$'s must contribute $-1$ charge each for $\phi_3$, and
similarly $-1$ for $\phi_4$ and $\phi_5$ each, and this can happen
in all possible combinations with appropriate antisymmetrization.

Let us denote by $u_{i,I}$ $(i=1,\cdots ,g-1)$, $I=3,4,5$, a
partition of the positions
of $T_F$'s which contribute $-1$ charge for $\phi_3$, $\phi_4$ and
$\phi_5$ respectively. Clearly, as a set $\{z_a\}=\bigcup_{i,I}
\{u_{i,I}\}$.
The contribution to the correlator from the internal sector is then:
\begin{eqnarray}
A_{g}^{(int)}&=&\frac{\prod_{I}\theta_{s,h_I}(\frac{1}{2}\sum_{i}(x_i+y_i)
-\sum_{i} u_{i,I})}{Z_{1}^{3/2}}
\frac{\prod_{I}\prod_{i<j} E(u_{i,I},u_{j,I})}
{\prod_{i,a}
E^{1/2}(x_i,z_a) E^{1/2}(y_i,z_a)}\nonumber\\ & &
\frac{\prod_{i<j}
E^{3/4}(x_i,x_j) E^{3/4}(y_i,y_j)}{\prod_{i,j} E^{3/4}(x_i,y_j)}
\prod_{i,I} \partial X_{I}(u_{i,I})~,
\label{int}
\end{eqnarray}
where $\theta_{s,h_I}$ are ${h_I}$-twisted $\theta_s$ functions.

d) From the determinants of the bosonic sector, $X_I$, $I=1,\cdots,5$,
as well as the $b,c$ ghost system.

Combining all the fermionic contributions, a), b) and c) we obtain:
\begin{eqnarray}
A_{g}^{(f)}&=&\frac{\theta^{2}_{s}(\frac{1}{2}\sum_{i}(x_i-y_i)+z-w)}
{\theta_{s}(\frac{1}{2}\sum_{i}(x_i+y_i)-\sum_{a}z_a+
2\Delta)}\frac
{\prod_{I}\theta_{s,h_I}(\frac{1}{2}\sum_{i}(x_i+y_i)-\sum_{i}
u_{i,I})}{Z_{1}^{2}}\nonumber\\ & &
\frac{\prod_{i<j} E(x_i,x_j) E(y_i,y_j)\prod_{I}E(u_{i,I},
u_{j,I})} {E^{2}(z,w)\prod_{a<b} E(z_a,z_b)}
\prod_{i}\frac{E(x_i,z)E(y_i,w)}{E(x_i,w)E(y_i,z)}\nonumber\\ & &
\frac{\prod_{i}\sigma(x_i)\sigma(y_i)}{\prod_{a}\sigma^{2}
(z_a)}\prod_{i,I}\partial X_{I}(u_{i,I})~.
\label{tot}
\end{eqnarray}
Notice that $\partial X_{I}$'s are replaced by zero modes, as they cannot
contract with each other.

The positions $z_a$ of the picture-changing operators are arbitrary.
To be able to do an explicit sum over spin structures we choose
the $z_a$'s to satisfy the condition:
\begin{equation}
\sum_{a=1}^{3g-3}z_{a}~=~\sum_{i=1}^{g-1}y_{i}-z+w+2\Delta~.
\label{gauge}
\end{equation}
For arbitrary $y_i$, $z$ and $w$ such $z_a$'s can always be chosen.
With this choice the $\theta$-function in the denominator of
(\ref{tot}), arising from superghosts, cancels with one $\theta$-
function in the numerator, arising from the space-time fermions.
Then the spin structure-dependent part of the amplitude is a
product of four $\theta$-functions:
\begin{equation}
\theta_{s}(\sum (x_i-y_i)/2+z-w)\prod_{I}\theta_{s,h_I}
(\sum (x_i+y_i)/2-\sum u_{i,I})~.\label{theta0}
\end{equation}
By considering the monodromy: $x_1\rightarrow x_1+a_i$ and
$x_1\rightarrow x_1+b_i$, one finds there are no relative phases
between different spin structures. We can then sum over all spin
structures using Riemann $\theta$-identity. The result is:
\begin{equation}
\theta(\sum x_{i}+z-w-\Delta)\theta_{-h_3}(\sum u_{i,3}-
\Delta)\theta_{-h_4}(\sum u_{i,4}-\Delta)
\theta_{-h_5}(\sum u_{i,5}-\Delta)~,\label{theta1}
\end{equation}
multiplied by the remaining terms in (\ref{tot}).

At this point, it is easy to see why the bosonic parts of the graviton
vertex operators, $\partial X^{\mu}
\bar{\partial} X^{\nu}$ do not contribute to the amplitude:
with appropriate choice of insertions, they give, after using
Riemann identity, an expression proportional to $\theta(\sum x_i
-\Delta)$ and this is zero by Riemann's vanishing theorem,
$\{\sum x_i\}$ being an effective divisor of degree $g\!-\!1$.

Now we use
the bosonization formula \cite{vv}:
\begin{equation}
\theta_{-h_I}(\sum_{i}u_{i}-\Delta)\prod_{i<j}E(u_i,u_j)
\prod_{i}\sigma(u_i)Z_{1}^{-1/2}~=~\det \omega_{-h_I,i}(u_j)
Z_{1,h_I}~.\label{bos}
\end{equation}
Here $\omega_{-h_I,i}$, with $i=1,\cdots,g-1$, are $(-h_I)$-twisted
holomorphic $1$-differentials and  $Z_{1,h_I}$ is the (chiral)
nonzero-mode determinant of the $h_I$-twisted $(1,0)$ system,
which cancels the corresponding bosonic contribution of d).
Using (\ref{bos}) for the three planes, we find that the total
amplitude becomes:
\begin{eqnarray}
& & \theta(\sum x_{i}+z-w-\Delta)\prod_{I}(\det\omega_{-h_I,i}
(u_{j,I}))\frac{\prod_{i<j}E(x_i,x_j)E(y_i,y_j)}{\prod_{a<b}
E(z_a,z_b)}\prod_{i}\frac{E(x_i,z)E(y_i,w)}{E(x_i,w)E(y_i,z)}
\nonumber\\ & & \frac{\prod_{i}\sigma(x_i)\sigma(y_i)}
{E^{2}(z,w)\prod_{a}
\sigma^{3}(z_a)}\prod_{i}[\partial X_{3}(u_{i,3})
\partial X_{4}(u_{i,4})\partial X_{5}(u_{i,5})]~,
\label{tot2}
\end{eqnarray}
times the bosonic contribution of the space-time $X$'s and
of the $(2,-1)$ $b,c$ system.

We further use the bosonization formula for the latter:
\begin{equation}
\theta(\sum z_a-3\Delta)\prod_{a<b}E(z_a,z_b)\prod\sigma^{3}(z_a)
Z_{1}^{-1/2}~=~\det h_{a}(z_b) Z_{2}~,\label{bc}
\end{equation}
where $Z_{2}$ is the nonzero-mode determinant of the $b-c$ system, and
$h_a$ with $a=1,\cdots,3g-3$ are the $3g\!-\!3$ quadratic differentials.
Similarly, the bosonization formula for the untwisted $(1,0)$ system is:
\begin{equation}
\theta(\sum x_{i}+z-w-\Delta)\frac{\prod_{i<j}E(x_i,x_j)}
{E(z,w)}
\prod_{i}\frac{E(x_i,z)}{E(x_i,w)}\frac{\sigma(z)\prod_{i}\sigma(x_i)}
{\sigma(w)}~=~\det\omega_{i}(x_j,z) Z_{1}~,
\label{1form}
\end{equation}
where $\omega_{i}$, $i=1,\cdots,g$, are $g$ abelian differentials.

Using further the fact that $\theta(\sum z_a-3\Delta) = \theta(\sum y_i
+z-w-\Delta)$ due to (\ref{gauge}), and once again using
the bosonization formula (\ref{1form}) with $x_i$ replaced by
$y_i$ and exchanging $z$ and $w$, we see that all the nonzero-mode
determinants cancel with the corresponding bosonic contributions,
with the result:
\begin{eqnarray}
A_{g}~&=&~\frac{1}{(\det Im\tau)^{2}} \det{\omega_{i}(x_j,z)}
\det{\omega_{i}(y_{j},w)} \frac{\prod_{I}\det{(\partial
X_{I}\omega_{-h_I,i} (u_{j,I}))}}{\det h_{a}(z_{b})}\nonumber\\ & &
\epsilon_{a_{1},\cdots,a_{3g-3}}\prod_{i=1}^{3g-3}
\int (\mu_{i}h_{a_i})
\,({\makebox{right-moving part}})\,({\makebox{lattice sum}})~,
\label{top}
\end{eqnarray}
where the lattice sum is due to the instanton contributions of
the $X_I$'s, $I=3,4,5$. The term containing $\int\mu h$,
$\mu_i$'s being Beltrami differentials, comes as usual due to the
necessity of soaking up the $3g\!-\!3$ $b$ zero modes and, once
multiplied by the right-moving part, gives the measure over
moduli space of genus $g$ Riemann surfaces.
$\tau$ is the period matrix, and the corresponding term in
(\ref{top}) comes from the integration over the space-time
$X^{\mu}$ zero-modes.

The result (\ref{top}) is for a fixed partition $\{u_{i,I}\}$
of the $z_a$'s. As mentioned earlier one must consider all
possible partitions and antisymmetrize. Furthermore, $\partial
X_{I}\omega_{-h_{I},i}$ are holomorphic quadratic differentials.
Therefore, summing aver all partitions $\{u_{i,I}\}$ with
the proper antisymmetrization gives:
\begin{equation}
\sum_{\{u\}}\prod_{I} \det(\partial X_{I}\omega_{-h_{I},i}
(u_{i,I}))~=~B \det h_{a}(z_b)~,\label{det}
\end{equation}
where $B$ is $z_a$-independent.

Moreover, taking into account the contribution from
the right-moving sector, which is $\bar{\partial}\bar{X}_{I}
\widetilde{\omega}_{h_{I},i}$ in the IIA case or $\bar{\partial}
{X}_{I}\widetilde{\omega}_{-h_{I},i}$ in the IIB case,
and integrating over $x_i,y_i,z,w$
one gets $(\det Im\tau)^2$ which cancels its inverse appearing in
(\ref{top}) times $(g!)^2$. Thus, the final result is:
\begin{equation}
A_{g}~=~(g!)^2 \int_{{\cal M}_g}B\bar{B} \det{\int({\mu_a}h_{b}})
\det{\int(\widetilde{\mu}_a \widetilde{h}_b)}
\,({\rm {lattice~sum}})~,\label{top1}
\end{equation}
where ${{\cal M}_g}$ is the moduli space of genus $g$ Riemann
surfaces.\footnote{Strictly speaking one descends on ${{\cal M}_g}$
after averaging over the orbifold group.}

Let us compare this result with the partition function $F_g$
of the (internal) topological field theory, obtained by
twisting the original $N=2$ SCFT corresponding
to the models A or B in the sense of \cite{Wit}:
\begin{equation}
F_{g}~=~\int_{{\cal M}_g} \prod_{a}\int ({\mu_a}G_{-})
\int ({\widetilde{\mu}_{a}}{\widetilde{G}}_{\pm}) e^{-S}~.\label{top2}
\end{equation}
Now $G_{-}=\sum_{I=3}^{5}\partial X_{I}{\bar{\psi}}_{I}$ and
$G_{+}=\bar{G}_{-}$, where ${\bar{\psi}}_{I}$
are dimension $1$ twisted fields,
having precisely $g\!-\!1$ zero-modes given by $\omega_{-h_I,i}$
, $i=1,\cdots,g-1$. In the right-moving sector the fermionic
$1$-forms are either $\widetilde{\psi}_{I}$ (A model), or
$\,\bar{\!\widetilde{\psi}}_{I}$ (B model) corresponding to the
two signs in (\ref{top2}).
All the nonzero-modes determinants cancel
out in (\ref{top2}) and only the classical instanton contributions
to $S$ survive, giving rise to a lattice sum. Using further the
definition of $B$ in (\ref{det}) one immediately gets that
$A_g~=~(g!)^2 F_g$.

So far we have considered a particular amplitude corresponding
to a particular kinematical configuration. All the other amplitudes
can be obtained by performing Lorentz transformations and $N=2$
space-time supersymmetry transformations. From two anti-self-dual Riemann
tensors and $2g\! -\!2$ anti-self-dual graviphoton field strengths one can
construct only three possible Lorentz-invariant combinations:
$R^2T^{2(g-1)}$, $(RT)^{2}T^{2(g-2)}$
and $(T\!RT)^{2}T^{2(g-3)}$
(for $g\ge 3$), where $(RT)^{2}=R_{\mu\nu\rho\sigma}T_{\rho\sigma}
R_{\mu\nu\alpha\beta}T_{\alpha\beta}$ and
$T\!RT= R_{\mu\nu\rho\sigma}
T_{\rho\sigma}T_{\mu\nu}$. However the last term implies a
non-vanishing result for an amplitude involving $g\!+\!1$
$S_1{\widetilde S}_1$ and $g\!-\!3$ $S_2{\widetilde S}_2$,
instead of $g\! -\!1$
each as in (\ref{ampl}). By following the same steps as above one can show
that this amplitude vanishes due to Riemann vanishing theorem. Thus, only
the first two Lorentz-invariant combinations arise in string computation.
In the Appendix we show that all such amplitudes
can be obtained from the following term in the effective action:
\begin{equation}
S_{\makebox{\scriptsize eff}}
{}~=~g F_{g}[R^2T^{2(g-1)}+2(g-1)(RT)^{2}T^{2(g-2)}]~.
\label{eff}
\end{equation}
\vglue 0.6cm
{\bf\noindent The hypermultiplet case}
\vglue 0.4cm
A similar analysis can be done for the dilaton-axion hypermultiplet whose
couplings depend only on the $(1,2)$ moduli which belong to
hypermultiplets. In fact, let us consider the $g$-th loop amplitude:
\begin{equation}
B_g = \langle (\partial\partial S)^2 (\partial Z)^{2g-2} \rangle_g\ ,
\label{Bampl}
\end{equation}
where $S$ is the complex dilaton-axion field and $Z$ is its complex R-R
partner belonging to the same $N=2$ hypermultiplet. Note that the
anti-self-duality condition in the gravitational multiplet case is now
replaced by chirality condition in terms of $N=1$ superfield language.
The vertex for $Z$ is given in (\ref{rs}), while $S$ involves those
for dilaton and axion in a specific combination. The dilaton vertex can be
obtained from (\ref{vg}) by taking a polarization tensor $h_{\mu\nu}$
proportional to $\delta_{\mu\nu}$. The axion vertex, however, needs
dualizing the vertex for the antisymmetric tensor. Fortunately, in the
following we will need only the fermion part of the vertex which comes
with two powers of momenta, and this allows one to go to the dual
description in a straightforward way. One finds that the vertex
$\partial\partial S$ in the 0 ghost picture contains the following
fermionic combinations: $\psi_1\psi_2$, ${\bar\psi}_1{\bar\psi}_2$ and
$\psi_1{\bar\psi}_1 +\psi_2{\bar\psi}_2$ in the left-moving part, together
with ${\widetilde\psi}_1{\,\bar{\!\widetilde\psi}}_2$,
${\,\bar{\!\widetilde\psi}}_1{\widetilde\psi}_2$ and
${\widetilde\psi}_1{\,\bar{\!\widetilde\psi}}_1 -
{\widetilde\psi}_2{\,\bar{\!\widetilde\psi}}_2$ in the right-moving part,
depending on the kinematics.

In particular, let us consider the kinematical configuration where $g\!-\!1$
of the $Z$'s appear with $S_1{\widetilde S}^{\dot{2}}$ and the remaining
$g\!-\!1$ with $S_2{\widetilde S}^{\dot{1}}$, where $S_{1,2}$ are defined in
(\ref{s1}, \ref{s2}) and
\begin{equation}
S^{\dot{1}} ~=~ {\rm exp}[\frac{i}{2}(\phi_1-\phi_2)]~,
\label{s1dot}\vspace{1mm}
\end{equation}
\begin{equation}
S^{\dot{2}} ~=~ {\rm exp}[-\frac{i}{2}(\phi_1-\phi_2)]~.
\label{s2dot}
\end{equation}
Furthermore, we take one $\partial\partial S$ vertex to be
$\psi_1\psi_2{\widetilde\psi}_1{\,\bar{\!\widetilde\psi}}_2$ and the second
one as ${\bar\psi}_1{\bar\psi}_2{\,\bar{\!\widetilde\psi}}_1
{\widetilde\psi}_2$. Following the same steps as in the computation of
the gravitational amplitude $A_g$, one finds that the result for the
left-moving part is identical with (\ref{top}), while for the
right-moving sector one gets a similar result except that
${\bar\partial}X_I{\bar\omega}_{h_I}$ is replaced by
${\bar X}_I{\bar\omega}_{-h_I}$ and the arguments $z$ and $w$ are
interchanged. The former can be understood as a result of the interchange
of $(1,1)$ and $(1,2)$ moduli, while the effect of the latter is that the
integrations of $x_i,y_i,z,w$ now give $g!(g-1)!(\det Im\tau)^2$.
Consequently, this amplitude is equal to $g!(g-1)!{\widetilde F}_g$,
where ${\widetilde F}_g$ is the topological partition function obtained
by twisting the right-moving sector in the opposite way as compared to the
case of $F_g$.

As in the gravitational case, we can obtain all the other kinematical
configurations by Lorentz and $N=2$ supersymmetry transformations. In the
Appendix, we show that they combine to the following term in the
effective action:
\begin{equation}
S_{\makebox{\scriptsize eff}}
{}~=~g {\widetilde F}_{g}[(\partial\partial S)^2
(\partial Z)^{2(g-1)} + 2(g-1)(\partial\partial S\partial Z)^{2}
(\partial Z)^{2(g-2)}]~,
\label{effhyper}
\end{equation}
where $(\partial\partial S)^2 =(\partial_{\mu}\partial_{\nu} S)
(\partial^{\mu}\partial^{\nu} S)$ and $(\partial\partial S\partial Z)^{2}
= (\partial_{\mu}\partial_{\nu} S\partial^{\nu} Z)
(\partial^{\mu}\partial_{\rho} S\partial^{\rho} Z)$.

\section{ The general Calabi-Yau case }

In the previous section, we saw that for orbifolds, since the internal
fermions are free, one could use Riemann theta identities to sum
the spin structures and thereby map the string amplitude to that of
topological theory. For the case of general Calabi-Yau spaces one cannot
of course follow this procedure. However the universal feature of all
the Calabi-Yau spaces is the underlying $N=2$ superconformal algebra and
in particular the $U(1)$ current algebra which can be bosonized in terms
of a free scalar field $H$ \cite{susy}. $H$ is a compact boson and
its momenta sit in a one-dimensional lattice given by the $U(1)$ charges of
the states. All the spin structure dependence enters in this
one-dimensional lattice by shifts as well as in the space-time fermions
and superghosts. The remaining part of the internal theory does not see
the spin structures. Therefore to do the spin structure sum it is
sufficient to know how it enters in the $U(1)$-charge lattice. On the
other hand the topological theory involves precisely twisting by adding
an appropriate background charge for the field $H$, and again the rest
of the internal theory is insensitive to this twisting. It is this
fortunate circumstance that will enable us to show the equivalence
between the string amplitude and $F_g$, without ever needing to know the
details of the Calabi-Yau space.

Let $\Gamma$ be the $U(1)$ lattice of $H$ momenta. The space-time fermions
define an $SO(2)\times SO(2)$ lattice. If one takes one of these $SO(2)$
lattices and combines with $\Gamma$, then it is known that the resulting
$2$-dimensional lattice is given by the coset $E_6/SO(8)$ \cite{Ler,LT}.
The characters are given by the branching functions $F_{\Lambda,s}(\tau)$
satisfying:
\begin{equation}
\chi_{\Lambda}(\tau) = \sum_s F_{\Lambda,s}(\tau)\chi_{s}(\tau)\ ,
\label{chils}
\end{equation}
where $\chi_{\Lambda}$ and $\chi_{s}$ are $E_6$ and $SO(8)$ level one
characters, $\Lambda$ denotes the three classes of $E_6$, and we are
using the spin structure basis denoted by $s$ to represent the four
conjugacy classes of $SO(8)$. The characters of the internal conformal
field theory times one complex space-time fermion can then be represented
as $ F_{\Lambda,s}(\tau){\rm Ch}_{\Lambda}(\tau)$, where ${\rm
Ch}_{\Lambda}(\tau)$ is the contribution of the rest of the internal
theory. The essential point here is that ${\rm Ch}_{\Lambda}(\tau)$
depends only on $\Lambda$ and not on the $SO(8)$ representations or
equivalently on the spin structures. Generalization of this to higher
genus is obtained by assigning an $E_6$ representation $\Lambda$ for
each loop and we will denote this collection by $\{ \Lambda\}$.

To proceed further, we note that in the amplitude (\ref{ampl}), due to the
conservation of $U(1)$ charge only $G_-$ and ${\widetilde G}_{\pm}$ parts
of the $T_F$ and ${\widetilde T}_F$ contribute, respectively, where $\pm$
correspond to type A and B models. As before, let us consider first the
contribution of the left-moving sector to the amplitude. The dependence of
$G_-$ on $H$ is given by:
\begin{equation}
G_- = e^{-iH/\sqrt{3}} {\hat G}_-\ ,
\label{gminus}
\end{equation}
where ${\hat G}_-$ has no singular operator product expansion with the
$U(1)$ current and it carries a dimension $4/3$. One can now explicitly
compute the correlation functions for the space-time fermions, superghost
and the free field $H$. After combining the lattice of (say) $\psi_2$
with that of $H$, the result can be expressed as:
\begin{eqnarray}
A_{g}^{(f)}&=&\frac{\theta_{s}(\frac{1}{2}\sum_{i}(x_i-y_i)+z-w)}
{\theta_{s}(\frac{1}{2}\sum_{i}(x_i+y_i)-\sum_{a}z_a+
2\Delta)} \nonumber\\ & &
\frac{F_{\{\Lambda\},s}(\frac{1}{2}\sum_{i}(x_i-y_i) +z-w\ ;\
\frac{\sqrt 3}{2}\sum_{i}(x_i+y_i)-\frac{1}{\sqrt 3}
\sum_{a}z_a)}{Z_{1}}\nonumber\\ & &
\frac{\prod_{i<j} E(x_i,x_j) E(y_i,y_j)}  {E^{2}(z,w)\prod_{a<b}
E^{2/3}(z_a,z_b)}
\prod_{i}\frac{E(x_i,z)E(y_i,w)}{E(x_i,w)E(y_i,z)}\nonumber\\ & &
\frac{\prod_{i}\sigma(x_i)\sigma(y_i)}{\prod_{a}\sigma^{2}
(z_a)}G_{ \{ \Lambda\} }(\{ z_a \} )~,
\label{acy}
\end{eqnarray}
where
\begin{equation}
G_{ \{ \Lambda\} }(\{ z_a \} ) = \langle \prod_a{\hat G}_-(z_a)
\rangle_{ \{ \Lambda\} }\ .
\label{glambda}
\end{equation}
$G_{ \{ \Lambda\} }$ represents the contribution of the internal conformal
field theory after removing the $H$ contribution and does not depend on
spin structure. $F_{\{\Lambda\},s}(u;v)$ represents the $SO(2)\times
\Gamma$ lattice contribution to genus g partition function with sources
$u$ and $v$ : $u$ coupled to $SO(2)$ lattice and $v$ coupled to $\Gamma$
\cite{Ler}.

To do the sum over spin structures we choose the positions $z_a$
satisfying (\ref{gauge}) and as a result the theta functions in the
numerator in the above equation cancels with that in the denominator. The
only spin structure dependence then appears in $F_{\{\Lambda\},s}$. We can
sum over the spin structures by using the formula \cite{Ler}:
\begin{equation}
\sum_s F_{\{ \Lambda\},s}(u;v) = F_{\{ \Lambda\}}(\frac{1}{2}u +
\frac{\sqrt{3}}{2}v\ ;\ \frac{\sqrt{3}}{2}u - \frac{1}{2}v) \ ,
\label{ssum}
\end{equation}
where
\begin{equation}
F_{\{\Lambda\}}(u;v) = \theta (u) \Theta_{\{\Lambda\}}(v) \ .
\label{flambda}
\end{equation}
$\Theta_{\{\Lambda\}}$ are given by:
\begin{equation}
\Theta_{\{\Lambda\}}(v) = \sum_{n_i \in {\bf Z}} exp\bigl(3 \pi i (n_i +
\frac{\lambda_i}{3})\tau_{ij} (n_j+\frac{\lambda_j}{3}) + 2\pi i\sqrt{3}
(n_i+\frac{\lambda_i}{3})v_i \bigr)\ ,
\label{theta2}
\end{equation}
where $i,j=1\cdots g$, and $\lambda_i = 0,1,2$ depending on $E_6$
conjugacy class $\Lambda_i$ corresponding to ${\bf 1}$, ${\bf{27}}$ and
${\overline{\bf {27}}}$, respectively. They can be expressed as combinations
of level six theta functions of \cite{Ler,LT}.
Using this formula, (\ref{acy}) becomes:
\begin{eqnarray}
A_{g}^{(f)}&=&\frac{\theta(\sum_{i}x_i+z-w)
\Theta_{\{\Lambda\}}(\frac{1}{\sqrt 3}
(\sum_{a}z_a-3\Delta))}{Z_{1}}\nonumber\\ & & \frac{\prod_{i<j} E(x_i,x_j)
E(y_i,y_j)}  {E^{2}(z,w)\prod_{a<b} E^{2/3}(z_a,z_b)}
\prod_{i}\frac{E(x_i,z)E(y_i,w)}{E(x_i,w)E(y_i,z)}\nonumber\\ & &
\frac{\prod_{i}\sigma(x_i)\sigma(y_i)}{\prod_{a}\sigma^{2} (z_a)}G_{ \{
\Lambda\} }(\{ z_a \} )~.
\label{acy1}
\end{eqnarray}

Going through now the same steps as in the case of orbifold, namely using
bosonisation formula for untwisted spin $(1,0)$ and $(2,-1)$ determinants
and using the condition (\ref{gauge}), we see that all the space-time
fermion boson as well as ghost and superghost non-zero mode determinants
cancel. After including the contribution of the right-moving sector, the
result is:
\begin{eqnarray}
A_{g}^{(f)}&=&\mid\prod_{a<b}E^{1/3}(z_a,z_b)}{\prod_{a}\sigma(z_a)
\frac{\Theta_{\{\Lambda\}}(\frac{1}{\sqrt 3} (\sum_{a}z_a-3\Delta)) G_{
\{\Lambda\} }(\{ z_a \} )}{\det h_a(z_b)} \nonumber\\ & & \frac{\det
\omega_i(x_j,z) \det \omega_i(y_j,w)}{\det Im\omega} \det \int \mu_a h_b
\mid^2~.
\label{acy2}
\end{eqnarray}
The $z_a$-dependent part in the numerator above, namely
\begin{equation}
B(\{ z_a \} ) \equiv \prod_{a<b}E^{1/3}(z_a,z_b) \prod_{a}\sigma(z_a)
\Theta_{\{\Lambda\}}(\frac{1}{\sqrt 3}
(\sum_{a}z_a-3\Delta)) G_{ \{\Lambda\} }(\{ z_a \} )\ ,
\label{bdef}
\end{equation}
transforms as quadratic differential in each $z_a$, is holomorphic with
first order zeroes as $z_a \rightarrow z_b$ for $a\ne b$ (generically)
and totally antisymmetric in $z_a$. This implies that
\begin{equation}
B(\{ z_a \} ) \propto \det h_a(z_b)\ .
\label{bdet}
\end{equation}
Integration over $x_i,y_i,z$ and $w$ gives $(g!)^2\cdot (\det Im\tau)^2$.
The final result is
\begin{equation}
A_g = (g!)^2 \int_{{\cal M}_g}\mid\det\int \mu_a (z_b) B(\{ z \} )\mid^2\ .
\label{afin}
\end{equation}

Now let us compare this result with the partition function of the
topological theory. Once again we extract the $H$ dependence in $G_-$
and explicitly work out its correlation function remembering that in the
twisted version there is a background charge $\frac{\sqrt 3}{2}\int
R^{(2)}H$ in the action. $3g\! -\! 3$ $G_-$`s appearing in the topological
partition function that are folded with the Beltrami differentials
precisely balance this background charge and one easily finds that
$F_g = A_g/(g!)^2$. This completes the proof that the string amplitude
under consideration is proportional to $F_g$ for a general Calabi-Yau
compactification.

\section{Effective field theory}

In this section we interpret the topological amplitude (\ref{eff})
and the holomorphic anomaly of Bershadsky et al.\ \cite{BCOV}
within the framework of the effective $N=2$ supergra\-vi\-ty theory
describing massless excitations of type II superstrings.

We shall follow the formalism of \cite{dWL} in which $N=2$ Poincar\'e
supergravity has been constructed
from conformal supergravity by imposing a set of gauge fixing constraints.
The anti-self-dual field strengths
$\widehat{F}^{I},~I=0,1,\dots,h$, of the vector
bosons, together with the corresponding scalars and fermions,
are contained in the ``reduced'' chiral multiplets $X^I$
of Weyl weights 1 \cite{chiraln2}:
\begin{equation}
X^I=X^I + \frac{1}{2}\widehat{F}^{I}_{\lambda\rho}
\epsilon_{ij} \theta^i\sigma_{\lambda\rho}\theta^j + \dots ,
\label{chir1}
\end{equation}
where the indices $i,j=1,2$ label the two supersymmetries.
In Poincar\'e supergravity, the scalar component of $X^0$ corresponds to
a  constrained field.
The unconstrained physical scalars of vector multiplets
-- the moduli -- are parametrized by
$Z^A\equiv X^A/X^0\, ,~A=1,\dots,h$. In type IIA theory, $h=h_{(1,1)}$,
whereas in type IIB, $h=h_{(1,2)}$  \nolinebreak\cite{CFG}. The
scalar component of $X^0$ can be expressed in terms of the
K\"ahler potential $K(Z,\bar{Z})$: $X^0=e^{K/2}$.

The amplitude (\ref{eff}) involves
graviphotons associated with the vertex operators (\ref{vgf}).
In order to relate
the ``superstring basis'' consisting of the graviphoton $T$ and
$h$ $U(1)$ gauge fields $F$ to the ``supergravity basis'' $\widehat{F}$
we consider the tree-level string amplitudes involving
two gauge bosons and one scalar, $Z$ or $\bar{Z}$.
The gauge kinetic terms are given by the matrix
$\N_{IJ}(Z,\bar{Z})=\frac{1}{4}\bar{\cal F}_{IJ}-(NX)_I(NX)_J/(X,NX)$,
where ${\cal F}(X)$ is the prepotential and the matrix $N$ is defined
as $N_{IJ}=\frac{1}{4}({\cal F}_{IJ}+\bar{\cal F}_{IJ})$ \cite{dWL}.
For anti-self-dual gauge fields, the three-point functions under
consideration  are given by the matrices $\langle\partial\bar{\N}/\partial
Z\rangle$ and $\langle\partial\bar{\N}/\partial \bar{Z}\rangle$
transformed to the superstring basis. On the other hand, the
$U(1)$ SCFT charge conservation
restricts non-zero couplings to $\bar{Z}FT$ and ${Z}FF$ only.
This is completely sufficient to conclude that, up to a normalization
factor,
\begin{equation}
T =\langle\,4 (N\bar{X})_I/(\bar{X},N\bar{X})\,
\rangle \widehat{F}^I.
\label{5t}
\end{equation}
This equation can be compared with the on-shell condition \cite{dWL}
for the auxiliary tensor $T^{ij}_{\mu\nu}$.
One finds that the superstring graviphoton
$T$ (\ref{5t}) is equal to the bosonic part of $\epsilon_{ij}T^{ij}$
evaluated in a constant scalar (moduli) background.
This is not surprising since the graviphoton vertex creates the
vector boson that enters
into the supersymmetry transformation of the gravitinos:
$\delta\psi^i_{\mu}=-\frac{1}{4}\sigma^{\lambda\rho}T_{\lambda\rho}^{ij}
\sigma_{\mu}\bar{\epsilon}_j+\dots$
The tensor $T_{\mu\nu}^{ij}$ belongs to the Weyl multiplet of conformal
supergravity.

After establishing connection between the graviphoton field strength
and the tensor component of the Weyl multiplet we proceed
to the construction of a $N=2$ supergravity Lagrangian
term containing the graviphoton-graviton interactions of eq.(\ref{eff}).
The Weyl multiplet is represented by the chiral superfield \cite{chiraln2}
\begin{equation}
W_{\mu\nu}^{ij}=T_{\mu\nu}^{ij}-
R_{\mu\nu\lambda\rho}\,\theta^i\sigma_{\lambda\rho}\theta^j+\dots,
\label{5w}
\end{equation}
which is anti-self-dual in its Lorentz indices and antisymmetric in
$i,j$. It has Weyl weight 1.
We define a scalar chiral superfield:
\begin{equation}
W^2 \equiv \epsilon_{ij}\epsilon_{kl}W_{\mu\nu}^{ij}W_{\mu\nu}^{kl}
=T_{\mu\nu}T_{\mu\nu}-2(\epsilon_{ij}\theta^i\sigma_{\mu\nu}
\theta^j)
R_{\mu\nu\lambda\rho}T_{\lambda\rho}-(\theta^i)^2(\theta^j)^2
R_{\mu\nu\lambda\rho}R_{\mu\nu\lambda\rho}+\dots
\label{5w2}\end{equation}

An invariant supergravity
action for chiral superfields can be constructed from the highest, i.e.\
$(\theta^i)^2(\theta^j)^2$, component of a scalar chiral superfield
of Weyl weight 2.
Let us consider an invariant action term
\begin{equation}
I_g=W^{2g}F_g(X) ,
\label{5ig}\end{equation}
where $F_g(X)$ is an analytic homogenous function of degree $2\!-\!2g$
of the superfields $X^I$.
It is easy to see that its highest component
contains the genus $g$ graviphoton-graviton interaction terms (\ref{eff}).
We conclude that $F_g W^{2g}$ provides a
supersymmetric completion of these interactions.
Note that the lowest component of $F_g$ can be written as
\begin{equation}
F_g(X)=(X^0)^{2-2g}F_g(Z)=e^{(1-g)K}F_g(Z)\, ,
\label{5f}\end{equation}
where $F_g(Z)$ is an analytic function of the moduli.

In the framework of effective supergravity theory,
the holomorphic anomaly of Bershadsky et al.\ amounts to non-analytic
moduli-dependence of functions $F_g$, as
a consequence of interactions that cannot be
represented by eq.(\ref{5ig}). This effect has been studied
before in \cite{AGN} at $g=1$.
As in the one-loop case, it is impossible to write down a standard
$N=2$ supergravity action term with non-holomorphic couplings of this type.
The reason is that anomalies are due to the propagation of massless
particles which leads to non-locality in the effective action.
An example of such a non-local term is
\begin{equation}
I_{g}^{\,\makebox{\scriptsize non-loc}}=W^{2g}\;{\Box}^{-2}
(\epsilon_{ij}\bar{\cal D}^i\bar{\sigma}_{\mu\nu}\bar{\cal D}^j)^2
(\epsilon^{kl}{\cal D}_k\sigma_{\lambda\rho}{\cal D}_l)^2 F_g(X,\bar{X}),
\label{5inl}
\end{equation}
where $\cal D$ and $\bar{\cal D}$ are supercovariant derivatives.
For a holomorphic $F_g$, $I_{g}^{\,\makebox{\scriptsize non-loc}}$
reduces to the local $I_g$ of eq.(\ref{5ig}).

According to Bershadsky et al., the moduli-dependence of the function
$F_g(Z,{\bar Z})$ is governed by the following equation:\footnote{Here again,
$F_g(Z,{\bar Z})\equiv e^{(g-1)K}F_g(X,\bar{X})$.}
\begin{equation}
\bar{\partial}_{\bar{A}}F_g=\bar{C}_{\bar{A}\bar{B}\bar{C}}e^{2K}
G^{B\bar{B}}G^{C\bar{C}}\left( D_BD_CF_{g-1}+\frac{1}{2}
\sum_rD_BF_r\cdot D_CF_{g-r}\right),
\label{5eq}
\end{equation}
where $G$ is the K\"ahler metric, $D$ are K\"ahler covariant
derivatives, and $C_{ABC}\equiv\frac{1}{4}X^0{\cal F}_{ABC}$. This
equation can be interpreted in the following way. As already mentioned,
the only non-vanishing tree-level couplings between one modulus and two
gauge bosons are of the form ${\bar Z}^{\bar A} F^{-B}T^-$ and
$Z^A F^{-B}F^{-C}$. The corresponding vertices can be normalized to
$G_{{\bar A}B}$ and $C_{ABC}$, respectively.
The first term of (\ref{5eq}) corresponds to the
diagrams of Fig.1, which involve genus $g\!-\!1$ vertices of the form
$$(g-1) (R^-T^-F^{-A}) (T^-)^{2(g-2)} Z^B D_A D_BF_{g-1}
\hspace{2.3cm}\makebox{(in Fig.1a),}$$
$$(g-1)(g-2) (R^-T^-)^2 (T^-)^{2(g-3)}Z^AZ^B D_A D_BF_{g-1}\,
\hspace{1.0cm}\makebox{(in Fig.1b),}$$
and tree-level vertices, giving rise to a non-holomorphic
contribution to the coupling
\linebreak
$(R^-T^-)^2 (T^-)^{2(g-2)}$.
A similar coupling, corresponding to the second term of (\ref{5eq}),
arises from the reducible diagram of Fig.2, involving
one genus $r$ vertex
$r (R^-T^-F^{-A}) (T^-)^{2(r-1)}$ $D_AF_r$, one
genus $g-r$ vertex of the same structure, and a tree-level vertex.
Thus genus $g$ non-holomorphicity is due to diagrams involving massless
particles interacting with lower genus interactions.

Let us briefly discuss here some aspects of spacetime
duality symmetries, like the small-large radius $SL(2,Z)$ symmetry
of orbifold compactifications and its Calabi-Yau analogues.
In the effective supergravity action, such symmetries are generated
by some special $N=2$ duality transformations which act on vector
multiplets \cite{dual}.
In particular, the transformations of scalar components $X^I$
are analytic. The graviton and auxiliary tensor $T^{ij}$ do not
transform  under $N=2$ duality
transformations, therefore $F_g(X)$ must be duality-invariant in order
to ensure invariance of $F_gW^{2g}$. However, one is usually interested
in the effective action written in terms of the
physical moduli $Z^I$ and their superpartners. $N=2$ duality induces
then K\"ahler transformations of the form $K\to K+\varphi+\bar{\varphi}$.

Although the constraints imposed in conformal supergravity in order
to obtain Poincar\'e theory are inert under duality transformations,
a specific superconformal $U(1)$ gauge choice like $X^0=\bar{X}^0=e^{K/2}$
is not duality-covariant. The unbroken symmmetry is then a superposition
of duality and a compensating, field-dependent $U(1)$ transformation.
After expressing the antisymmetric tensor of eq.(\ref{5t})
in terms of the physical moduli and gauge bosons one finds that
it transforms as $T\to e^{(\bar{\varphi} -\varphi )/2}T$.
$F_g(Z)$ transforms as $F_g\to e^{(2g-2)\varphi}F_g$. For example,
in orbifold compactifications $F_g$ transforms under $SL(2,Z)$ symmetry
associated with each plane as a modular form of weight $g\!-\!1$.

\setcounter{equation}{0}

\section{Conclusions}

In this paper we have shown that the topological partition function $F_g$
appears as the coupling of the term $W^{2g}$,
where $W$ is the chiral superfield of
$N=2$ super\-gra\-vi\-ta\-tional mul\-ti\-plet, in the $g$-th loop,
four-dimensional type II string effective action. The
holomorphic anomaly of $F_g$ is related to the non-locality of the
effective action due to the propagation of massless states. We also
obtained the corresponding term in the effective action for dilaton
hypermultiplets in component form. In this case $F_g$ is replaced by
${\widetilde F}_g$ which is the partition function of the topological
theory obtained by twisting the right movers in the opposite way. It
would be interesting to express this effective action in the superfield
language.

The fact that $F_g$'s play the role of chiral couplings in the
effective $N=2$ supergravity theory, whose violation of holomorphicity is
related to anomalous field theory graphs involving propagation of
massless states, suggests that they are subject to non-renormalization
theorems. This property was indeed proven for the moduli dependence of
$F_1$ \cite{ANT} which, in the case of type II strings, coincides with
the gravitational $R^2$ couplings. Moreover, from the analysis of
Section 3 it is straightforward to show that the coupling of $W^{2n}$
vanishes for all genera $g<n$. An additional indication that the
term $W^{2n}$ appears only at genus $g=n$ has to do with the dilaton
which plays the role of the string loop expansion parameter, while
simultaneously in type II strings it belongs to a $N=2$ hypermultiplet.
Non zero corrections to the coupling of $W^{2n}$ for $g>n$ would mix the
dilaton with vector multiplets which is forbidden in $N=2$
supergravity.

An important consequence of such non-renormalization theorem is that the
anti-self-dual part of the effective action ($W^+=0$) is completely
determined by the topological partition function, to all orders in string
perturbation theory. Moreover, if $W^+=0$ is consistent with the string
equations of motion, this part of the theory is sufficient to generate
all anti-self-dual solutions. The recursion relation satisfied by $F_g$
can then be used to determine the moduli dependence of the corresponding
couplings from tree-level quantities, while the corresponding master
equation could be used to sum up the perturbative series and obtain
non-perturbative information. Unfortunately, the recursion relation does
not determine $F_g$'s uniquely. Target space modular invariance and
asymptotic behavior are not a priori sufficient to fix the holomorphic
ambiguity of the homogeneous solution which is a modular form of weight
$2g\! -\!2$. If one were able to solve the ambiguity problem and sum up the
series, one could in principle apply the result to understand some
non-perturbative features of string theory, such as the generation of
non-perturbative moduli potential or the problem of supersymmetry
breaking.

Another important open question is the generalization of these results
in the case of heterotic string which is of more physical interest.
\\[1cm]
{\bf Acknowledgements}

We thank S. Cecotti, S. Ferrara, H. Ooguri
and particularly C. Vafa for valuable discussions.
K.S.N. and T.R.T. acknowledge the hospitality of the Centre de Physique
Th\'eorique at Ecole Polytechnique, during completion of this work.

\begin{flushleft}
{\large\bf Appendix}\end{flushleft}

In this Appendix, we obtain the effective action terms corresponding
to the amplitudes computed in Sections 3 and 4.
\renewcommand{\theequation}{A.\arabic{equation}}
\renewcommand{\thesection}{A.}
\setcounter{equation}{0}
\vglue 0.3cm
{\bf\noindent The gravitational case}
\vglue 0.2cm

Let us first work out the helicity components appearing in the
anti-self-dual part of the various vertex operators defined in Section 2.
Following Section 3, the four space-time fermionic coordinates
$\Psi^{\mu}$ form two complex fermions $\psi_{1,2}$,
\begin{equation}
\psi_1=\frac{\Psi^1-i\Psi^2}{\sqrt 2}\ ,\ \ \ \ \ \ \ \
\psi_2=\frac{\Psi^0-i\Psi^3}{\sqrt 2} \ ,
\label{psi12}
\end{equation}
where $\Psi^{1,2}$ and $\Psi^3$ denote the transverse and longitudinal
components, respectively, and they are bosonized in terms of $\phi_{1,2}$
according to (\ref{boso}), (\ref{s1}) and (\ref{s2}).

The anti-self-duality
condition for the graviphoton (\ref{vgf}) implies that its polarization
vector $\epsilon_{\mu}$ satisfies the condition ${\bar\sigma}^{\mu\nu}
\epsilon_{\mu}p_{\nu}=0$ which gives:
\begin{equation}
\epsilon_i p_0=\varepsilon_{ijk}\epsilon_j p_k\ ,
\label{esd}
\end{equation}
in the gauge $\epsilon_0=0$. Using (\ref{esd}), one finds for the three
independent anti-self-dual components of the field strength $T$ the
following space-time part of the graviphoton vertex:
\begin{equation}
T_{0+}S_1{\widetilde S}_1 - T_{0-}S_2{\widetilde S}_2 -
T_{03}(S_1{\widetilde S}_2 + S_2{\widetilde S}_1)\ ,
\label{dgv}
\end{equation}
where $T_{0\pm}=T_{01}\pm iT_{02}$.

The anti-self-duality condition for the graviton (\ref{vg}) in the gauge
$h_{00}=1, h_{0i}=0$ implies that $h_{ij}=\epsilon_i\epsilon_j$ where
$\epsilon_i$ satisfies (\ref{esd}). As a result, the fermionic part of
the graviton vertex factorizes in left and right-moving parts and after a
simple algebra one finds that the five independent anti-self-dual
components of the Riemann tensor are:
\begin{eqnarray}
& &R_{0+0+}\psi_1\psi_2{\widetilde\psi}_1{\widetilde\psi}_2 +
R_{0-0-}{\bar\psi}_1{\bar\psi}_2
{\,\bar{\!\widetilde\psi}}_1{\,\bar{\!\widetilde\psi}}_2 + \nonumber\\ & &
R_{0303}[\psi_1\psi_2{\,\bar{\!\widetilde\psi}}_1{\,\bar{\!\widetilde\psi}}_2 +
{\bar\psi}_1{\bar\psi}_2{\widetilde\psi}_1{\widetilde\psi}_2 +
(\psi_1{\bar\psi}_1 + \psi_2{\bar\psi}_2)
({\widetilde\psi}_1{\,\bar{\!\widetilde\psi}}_1 +
{\widetilde\psi}_2{\,\bar{\!\widetilde\psi}}_2)] + \nonumber\\ & &
R_{030+}[\psi_1\psi_2 ({\widetilde\psi}_1{\,\bar{\!\widetilde\psi}}_1 +
{\widetilde\psi}_2{\,\bar{\!\widetilde\psi}}_2) +
(\psi_1{\bar\psi}_1 + \psi_2{\bar\psi}_2)
{\widetilde\psi}_1{\widetilde\psi}_2] + \nonumber\\ & &
R_{030-}[{\bar\psi}_1{\bar\psi}_2
({\widetilde\psi}_1{\,\bar{\!\widetilde\psi}}_1 +
{\widetilde\psi}_2{\,\bar{\!\widetilde\psi}}_2) +
(\psi_1{\bar\psi}_1 + \psi_2{\bar\psi}_2)
{\,\bar{\!\widetilde\psi}}_1{\,\bar{\!\widetilde\psi}}_2]\ .
\label{dg}
\end{eqnarray}

\hspace*{-3mm}The amplitude $A_g$ computed in Section 3 corresponds to the term
$R_{0+0+}R_{0-0-}(T_{0+}T_{0-})^{g-1}$
which appears in both terms of the
effective action (\ref{eff}). Adding both contributions, one finds that it
comes with a coefficient $g^2F_g$. Taking into account the combinatoric
factor $(g-1)!^2$, the amplitude $A_g$ should be $(g!)^2F_g$, which is
indeed the result of the string computation found in Section 3. To
determine the relative coefficients between the two terms in (\ref{eff}),
we consider also the term $R_{0-0-}R_{0303}(T_{0+})^g(T_{0-})^{g-2}$. This
appears only in the second term of the effective action and comes with a
coefficient $-g(g-1)F_g$. Taking into acount the combinatorial factor
$g!(g-2)!$, the corresponding amplitude should be $-(g!)^2F_g$.

To prove this, we start with $A_g$ in (\ref{ampl}) and choose one of the
$T_{0-}$ vertices to convert it into a graviton vertex by a space-time
supersymmetry transformation. In fact substituting the following identity,
\begin{eqnarray}
e^{-1/2(\phi+\widetilde{\phi})} S_2{\widetilde S}_2\Sigma(y) &=&
\oint dz e^{-\frac{1}{2}\phi+i\frac{\sqrt 3}{2}H} S_2 (z)
\oint d\zbar e^{-\frac{1}{2}\widetilde{\phi}\mp i\frac{\sqrt 3}{2}H}
{\widetilde S}_2 (\zbar)\nonumber\\ & &
(\psi_1{\bar\psi}_1 + \psi_2{\bar\psi}_2)
({\widetilde\psi}_1{\,\bar{\!\widetilde\psi}}_1 +
{\widetilde\psi}_2{\,\bar{\!\widetilde\psi}}_2) (y)\ ,
\label{sst}
\end{eqnarray}
in (\ref{ampl}) and deforming the contours, one finds that the only
non-zero contribution comes when both contours encircle the graviton
vertex from $R_{0+0+}$, namely
$\psi_1\psi_2{\widetilde\psi}_1{\widetilde\psi}_2$. As a result, this
graviton is converted into $e^{-1/2(\phi+\widetilde{\phi})} S_1{\widetilde
S}_1\Sigma$ which corresponds to $T_{0+}$, while one of the $T_{0-}$ was
converted into $R_{0303}$, as shown by the expression (\ref{sst}). Thus,
the amplitude corresponding to $R_{0-0-}R_{0303}T_{0+}^gT_{0-}^{g-2}$ is
equal to $-A_g$, where the minus sign comes from the relative sign
between $T_{0+}$ and $T_{0-}$ in (\ref{dgv}). This completes the form of
the effective action (\ref{eff}).
\vglue 0.3cm
{\bf\noindent The hypermultiplet case}
\vglue 0.2cm

{}From (\ref{rs}) we see that the vertex for the R-R scalar $\partial Z$ has
the following kinematic combinations:
\begin{equation}
\partial_+ Z S_1{\widetilde S}^{\dot 1} -
\partial_- Z S_2{\widetilde S}^{\dot 2} -
\partial_l Z S_1{\widetilde S}^{\dot 2} -
\partial_{\bar l} Z S_2{\widetilde S}^{\dot 1}\ ,
\label{dz}
\end{equation}
where $\partial_{\pm}=\partial_1\pm\partial_2$,
$\partial_l=\partial_0+\partial_3$, and $S_{1,2}$, $S_{{\dot 1},{\dot 2}}$
are given in (\ref{s1}-\ref{s2}), (\ref{s1dot}-\ref{s2dot}). To obtain
the vertex operator for $\partial\partial S$, we need to consider the
vertex for the dilaton $D$ and axion $b$. The fermionic part of the
dilaton vertex is simply given by:
\begin{equation}
\partial_{\alpha}\partial_{\beta}D \psi_{\alpha}\psi_{\gamma}
{\widetilde\psi}_{\beta}{\widetilde\psi}_{\gamma}\ .
\label{dil}
\end{equation}
As for the axion vertex we use the defining relation
\begin{equation}
\partial_{\mu} b= \frac{1}{6}\varepsilon_{\mu\nu\lambda\sigma}
\partial_{\nu}B_{\lambda\sigma}\ ,
\label{defb}
\end{equation}
and the fermionic part of the vertex for the antisymmetric tensor
$B_{\lambda\sigma} \psi_{\mu}\psi_{\lambda}{\widetilde\psi}_{\nu}
{\widetilde\psi}_{\sigma}$. After a straightforward algebra one can show
that the axion vertex is given by:
\begin{equation}
\partial_{\alpha}\partial_{\beta}b\,
\varepsilon_{\alpha\mu\nu\rho} \psi_{\mu}\psi_{\nu}
{\widetilde\psi}_{\beta}{\widetilde\psi}_{\rho}\ .
\label{defax}
\end{equation}
By using the relation $S=D+ib$, one can then obtain the vertex for
$\partial\partial S$, which contains nine independent components arising
from taking all combinations of $\psi_1\psi_2$, ${\bar\psi}_1
{\bar\psi}_2$, $(\psi_1{\bar\psi}_1 + \psi_2{\bar\psi}_2)$ from the left,
and ${\widetilde\psi}_1 {\,\bar{\!\widetilde\psi}}_2$,
${\,\bar{\!\widetilde\psi}}_1 {\widetilde\psi}_2$ and $({\widetilde\psi}_1
{\,\bar{\!\widetilde\psi}}_1 - {\widetilde\psi}_2
{\,\bar{\!\widetilde\psi}}_2)$
from the right. In the following we show only the terms that we will need
here:
\begin{equation}
\partial_+\partial_+S \psi_1\psi_2
{\widetilde\psi}_1{\,\bar{\!\widetilde\psi}}_2 +
\partial_-\partial_-S {\bar\psi}_1{\bar\psi}_2
{\,\bar{\!\widetilde\psi}}_1 {\widetilde\psi}_2 +
\partial_+\partial_l S \psi_1\psi_2 ({\widetilde\psi}_1
{\,\bar{\!\widetilde\psi}}_1 - {\widetilde\psi}_2 {\,\bar{\!\widetilde\psi}}_2)
+ \cdots
\label{dds}
\end{equation}

The kinematic configuration of the amplitude (\ref{Bampl}) considered in
Section 3 corresponds to $\partial_+\partial_+S \partial_-\partial_-S
(\partial_l Z)^{g-1} (\partial_{\bar l} Z)^{g-1}$, which appears only in
the first term of the effective action (\ref{effhyper}) and it comes with
a coefficient $g{\widetilde F}_{g}$. Taking into account the
combinatorial factor $(g-1)!^2$ one finds that the corresponding
amplitude should be $g!(g-1)!{\widetilde F}_{g}$, which was indeed the
result found in Section 3. To determine the coefficient of the second
term in the effective action, we consider $\partial_-\partial_- S
\partial_+\partial_l S \partial_+ Z (\partial_l Z)^{g-2}
(\partial_{\bar l} Z)^{g-1}$. It is easy to see that this appears only in
the second term of (\ref{effhyper}) and comes with the coefficient
\linebreak
$g(g-1){\widetilde F}_{g}$. Taking into account the combinatorial factor
$(g-2)!(g-1)!$ one should expect the corresponding amplitude to be
$g!(g-1)!{\widetilde F}_{g}$. We now show that this is indeed the result
of string computation.

Let us start from the amplitude computed in Section 3, namely
$\partial_+\partial_+S \partial_-\partial_-S
(\partial_l Z)^{g-1}$ $(\partial_{\bar l} Z)^{g-1}$. Choose one of the
$\partial_l Z$ vertices, $e^{{-1/2}(\phi+\widetilde\phi)}
S_1{\widetilde S}^{\dot{2}}\Phi$, and express it as supersymmetric
transformation of the $\partial_+\partial_l S$ vertex (\ref{dds}) as
follows:
\begin{eqnarray}
e^{{-1/2}(\phi+\widetilde\phi)} S_1{\widetilde S}^{\dot{2}}\Phi(y) &=&
\oint dz e^{-\frac{1}{2}\phi+i\frac{\sqrt 3}{2}H} S_2 (z)
\oint d\zbar e^{-\frac{1}{2}\widetilde{\phi}\pm i\frac{\sqrt 3}{2}H}
{\widetilde S}^{\dot{2}} (\zbar)\nonumber\\ & & \psi_1\psi_2
({\widetilde\psi}_1{\,\bar{\!\widetilde\psi}}_1 -
{\widetilde\psi}_2{\,\bar{\!\widetilde\psi}}_2) (y)\ .
\label{ddssusy}
\end{eqnarray}
Once again, upon deforming the two contours, one finds that the only
non-vanishing contribution comes when both the contours encircle the
vertex $\psi_1\psi_2{\widetilde\psi}_1{\,\bar{\!\widetilde\psi}}_2$ of
$\partial_+\partial_+ S$, converting it into the vertex
$e^{{-1/2}(\phi+\widetilde\phi)} S_1{\widetilde S}^{\dot{1}}\Phi$ of
$\partial_+ Z$. This is exactly the amplitude we wanted to compute, and
the above analysis shows that it is equal to the amplitude computed in
Section 3, namely $g!(g-1)!{\widetilde F}_{g}$, which is the result we
wanted to establish.

Finally, the only other Lorentz-invariant combination (for $g\ge 3$) is
$(\partial\partial S\partial Z\partial Z)^2$
$(\partial Z)^{2(g-3)}$. This
would contain a term $(\partial_-\partial_-S)^2 (\partial_+ Z)^{g+1}
(\partial_- Z)^{g-3}$, which can be shown to vanish after summing over
the spin structures due to Riemann vanishing theorem. This proves that
the string computation indeed gives rise to the effective action
(\ref{effhyper}).

\newpage
\begin{figcap}
\item Field-theoretical diagrams contributing to the first term in the
holomorphic anomaly equation (\ref{5eq}). $h$ is the graviton.
\item Field-theoretical diagram contributing to the second term in the
holomorphic anomaly equation (\ref{5eq}).
\end{figcap}

\newpage
\begin{thebibliography}{99}
\bibitem{DKL} L.J. Dixon, V.S. Kaplunovsky and J. Louis,
\np 355 (1991) 649.
\bibitem{AGN} I. Antoniadis, E. Gava and K.S. Narain, \np 383 (1992) 93.
\bibitem{CFIV} S. Cecotti, P. Fendley, K. Intriligator and C. Vafa,
\np 386 (1992) 405.
\bibitem{ft} J.-P. Derendinger, S. Ferrara, C. Kounnas and F.
Zwirner, \np 372 (1992) 145;
G.L. Cardoso and B.A. Ovrut, \np 369 (1992) 351.
\bibitem{AGNT} I. Antoniadis, E. Gava and K.S. Narain and T.R. Taylor,
preprint NUB-3057 (1992),
hep-th/9212045, to appear in Nucl.\ Phys.\ B.
\bibitem{BCOV} M. Bershadsky, S. Cecotti, H. Ooguri and C. Vafa,
preprint HUTP-93/A008 (1993), hep-th/9302103.
\bibitem{Top} E. Witten, \cm 118 (1988) 411;
T. Eguchi and S.-K. Yang, Mod. Phys. Lett. A4 (1990) 1653.
\bibitem{bp} C. Vafa, private communication.
\bibitem{chiraln2} M. de Roo, J.W. Van Holten, B. de Wit and A. Van
Proeyen, \np 173 (1980) 175; E. Bergshoeff, M. de Roo and B. de Wit,
\np 182 (1981) 173.
\bibitem{bf} M. Bershadsky, S. Cecotti, H. Ooguri and C. Vafa, to
appear.
\bibitem{Sei} N. Seiberg, \np 303 (1988) 286.
\bibitem{CFG} S. Cecotti, S. Ferrara and L. Girardello, Int.
Journal of Mod. Phys. A4 (1989) 2475.
\bibitem{DKV} L. Dixon, V. Kaplunovsky and C. Vafa, \np 294
(1987) 43.
\bibitem{dWL} B. de Wit, P.O. Lauwers and A. van Proeyen, \np 255 (1985) 569;
E. Cremmer, C. Kounnas, A. Van Proeyen, J.P. Derendinger, S. Ferrara,
B. de Wit and L. Girardello, \np 250 (1985) 385.
\bibitem{FMS} D. Friedan, E. Martinec, S. Shenker, \np 271 (1986) 93.
\bibitem{vv} E. Verlinde and H. Verlinde, \np 288 (1987) 357.
\bibitem{Wit} E. Witten, ``Mirror Manifolds and Topological Field
Theory", in {\it Essays on Mirror Manifolds}, ed. S.-T. Yau,
(International Press,1992) p.37.
\bibitem{susy} W. Boucher, D. Friedan and A. Kent, \pl 172 (1986)
316; A. Sen, \np 278 (1986) 289;
T. Banks, L. Dixon, D. Friedan and E. Martinec, \np 299 (1987) 613.
\bibitem{Ler} W. Lerche, A.N. Schellekens and N.P. Warner,
Phys. Rep. 117 (1989) 1;
O. Lechtenfeld and W. Lerche  \pl 227 (1989) 373.
\bibitem{LT} D. L\"{u}st and S. Theisen, \pl 227 (1989) 367.
\bibitem{dual} B. de Wit and A. van Proeyen, \np 245 (1984) 89;
S. Ferrara, D. L\"ust and S. Theisen, \pl 242 (1990) 39.
\bibitem{ANT} I. Antoniadis, K.S. Narain and T.R. Taylor, \pl 267
(1991) 37.

\end{thebibliography}
\end{document}